\documentclass[acmsmall]{acmart}

\usepackage[leqno]{amsmath}
\usepackage{multicol}
\usepackage{listings}
\usepackage{fancyvrb}

\AtBeginDocument{%
  \providecommand\BibTeX{{%
    \normalfont B\kern-0.5em{\scshape i\kern-0.25em b}\kern-0.8em\TeX}}}
\setcopyright{acmcopyright}
\copyrightyear{2021}
\acmYear{2021}
\acmDOI{10.1145/1122445.1122456}

\acmJournal{JACM}
\acmVolume{39}
\acmNumber{4}
\acmArticle{111}
\acmMonth{8}

\makeatletter
\newcommand{\leqnomode}{\tagsleft@true\let\veqno\@@leqno}
\newcommand{\reqnomode}{\tagsleft@false\let\veqno\@@eqno}
\makeatother
\begin{document}

\title{So You Want to Analyze Scheme Programs With Datalog?}

\author{Davis Ross Silverman}
\authornote{Both authors contributed equally to this research.}
\email{dasilver@syr.edu}
\orcid{0000-0002-9921-0176}
\author{Yihao Sun}
\authornotemark[1]
\email{ysun67@syr.edu}
\affiliation{%
  \institution{Syracuse University}
  \streetaddress{900 S Crouse Ave}
  \city{Syracuse}
  \state{New York}
  \country{USA}
  \postcode{13244}
}
\orcid{0000-0003-0946-2511}

\author{Kristopher Micinski}
\affiliation{%
  \institution{Syracuse University}
  \streetaddress{900 S Crouse Ave}
  \city{Syracuse}
  \state{New York}
  \country{USA}
  \postcode{13244}
}
\email{kkmicins@syr.edu}

\author{Thomas Gilray}
\affiliation{%
  \institution{University of Alabama at Birmingham}
  \streetaddress{1720 University Blvd}
  \city{Birmingham}
  \state{Alabama}
  \country{USA}
  \postcode{35294}
}
\email{gilray@uab.edu}

\renewcommand{\shortauthors}{Silverman and Sun, et al.}

\newcommand\mdoubleplus{\ensuremath{\mathbin{+\mkern-9mu+}}}
\newcommand\mae{\ensuremath{\text{\ae}}}
\newcommand{\lamsyn}[2]{(\lambda\;(#1 ...)\;#2)}
\newcommand{\singlelamsyn}[2]{(\lambda\;(#1)\;#2)}
\newcommand{\vararglamsyn}[2]{(\lambda\;#1\;#2)}
\newcommand{\letsyn}[3]{(\texttt{let}\;((#1\;#2)\;...)\;#3)}
\newcommand{\ifsyn}[3]{(\texttt{if}\;#1\;#2\;#3)}
\newcommand{\primsyn}[2]{(\texttt{prim}\;#1\;#2\;...)}
\newcommand{\singleprimsyn}[1]{(\texttt{prim}\;#1)}
\newcommand{\applyprimsyn}[2]{(\texttt{apply-prim}\;#1\;#2)}
\newcommand{\applysyn}[2]{(\texttt{apply}\;#1\;#2)}
\newcommand{\callccsyn}[1]{(\texttt{call/cc}\;#1)}
\newcommand{\setsyn}[2]{(\texttt{set!}\;#1\;#2)}
\newcommand{\quotesyn}[1]{(\texttt{quote}\;#1)}
\newcommand{\aeval}[0]{\widehat{\mathcal{A}}}

\newcommand{\truesyn}{\textbf{\texttt{\#t}}}
\newcommand{\falsesyn}{\textbf{\texttt{\#f}}}

\newcommand{\Call}[0]{\textsf{Call}}
\newcommand{\Let}[0]{\textsf{Let}}
\newcommand{\llet}[0]{\textit{let}} 
\newcommand{\Lam}[0]{\textsf{Lam}}
\newcommand{\lam}[0]{\textit{lam}}
\newcommand{\Exp}[0]{\textsf{Exp}}
\newcommand{\Var}[0]{\textsf{Var}}
\newcommand{\sstate}[0]{\hat{\varsigma}} 
\newcommand{\State}[0]{\hat{\Sigma}}
\newcommand{\Eval}[0]{\widehat{\textit{Eval}}}
\newcommand{\Apply}[0]{\widehat{\textit{Apply}}}
\newcommand{\Env}[0]{\widehat{\textit{Env}}}
\newcommand{\Store}[0]{\widehat{\textit{Store}}}
\newcommand{\KStore}[0]{\widehat{\textit{KStore}}}
\newcommand{\VStore}[0]{\widehat{\textit{VStore}}}
\newcommand{\KAddr}[0]{\widehat{\textit{KAddr}}}
\newcommand{\vaddr}[0]{\hat{a}_{v}}
\newcommand{\kaddr}[0]{\hat{a}_{\hat{\kappa}}}
\newcommand{\Val}[0]{\widehat{\textit{Val}}}
\newcommand{\env}[0]{\hat{\rho}}
\newcommand{\store}[0]{\hat{\sigma}}
\newcommand{\kstore}[0]{\store_{\kont}}
\newcommand{\vstore}[0]{\store_{\val}}
\newcommand{\VAddr}[0]{\widehat{\textit{VAddr}}}
\newcommand{\Kont}[0]{\widehat{\textit{Kont}}}
\newcommand{\val}[0]{\hat{v}}
\newcommand{\closure}[0]{\widehat{clo}}
\newcommand{\Closure}[0]{\widehat{\textit{Clo}}}
\newcommand{\kont}[0]{\hat{\kappa}}
\newcommand{\done}[0]{\widehat{done}}
\newcommand{\IVal}[0]{\widehat{\textit{InnerVal}}}
\newcommand{\ival}[0]{\hat{iv}}
\newcommand{\CALL}[0]{\widehat{CALL}}
\newcommand{\mtk}[0]{\widehat{\textbf{mtk}}}
\newcommand{\ifk}[4]{\widehat{\textbf{ifk}}(#1 , #2 , #3 , #4)}
\newcommand{\setk}[2]{\widehat{\textbf{setk}}(#1 , #2)}
\newcommand{\callcck}[2]{\widehat{\textbf{callcck}}(#1 , #2)}
\newcommand{\letk}[4]{\widehat{\textbf{let}}(#1 , #2 , #3 , #4)}
\newcommand{\fnk}[4]{\widehat{\textbf{fn}}(#1 , #2 , #3 , #4)}
\newcommand{\argk}[4]{\widehat{\textbf{arg}}(#1 , #2 , #3 , #4)}
\newcommand{\primsk}[4]{\widehat{\textbf{p1}}(#1 , #2 , #3 , #4)}
\newcommand{\primfk}[3]{\widehat{\textbf{p2}}(#1 , #2 , #3)}
\newcommand{\TRUTHY}[0]{\widehat{TRUTHY}}
\newcommand{\FALSY}[0]{\widehat{FALSY}}
\newcommand{\new}[0]{\widehat{new}}
\newcommand{\allock}[0]{\widehat{alloc_k}}
\newcommand{\allocv}[0]{\widehat{alloc_v}}
\newcommand{\Context}[0]{\widehat{\textit{Context}}}
\newcommand{\context}[0]{\widehat{ctx}}
\newcommand{\Prim}[0]{\textsf{Prim}}
\newcommand{\prim}[0]{\textit{op}}
\newcommand{\PrimPoint}[0]{\widehat{\textit{PVal}}}
\newcommand{\primPoint}[0]{\widehat{\textit{p}}}
\newcommand{\E}[4]{E\langle #1 , #2 , #3 , #4 \rangle}
\newcommand{\A}[3]{A\langle #1 , #2 , #3 \rangle}
\newcommand{\inj}[0]{\hat{\mathcal{I}}}
\newcommand{\copyctx}[0]{\widehat{\textit{copy}}}
\newcommand{\stateSpace}[0]{\hat{\xi}}
\newcommand{\StateSpace}[0]{\hat{\Xi}}
\newcommand{\reachableConfig}[0]{\hat{r}}
\newcommand{\ReachableConfig}[0]{\hat{R}}
\newcommand{\config}[0]{\hat{c}}
\newcommand{\Config}[0]{\hat{C}}
\newcommand{\ConfigE}[3]{CE\langle #1 , #2 , #3 \rangle}
\newcommand{\ConfigA}[2]{CA\langle #1 , #2 \rangle}

\begin{abstract}
Static analysis approximates the results of a program by examining
only its syntax. For example, control-flow analysis (CFA) determines
which syntactic lambdas (for functional languages) or (for
object-oriented) methods may be invoked at each call site within a
program. Rich theoretical results exist studying control flow analysis
for Scheme-like languages, but implementations are often complex and
specialized. By contrast, object-oriented languages (Java in
particular) enjoy high-precision control-flow analyses that scale to
thousands (or more) of lines of code. State-of-the-art implementations
(such as DOOP on Souffl\'e) structure the analysis using Horn-SAT
(Datalog) to enable compilation of the analysis to efficient
implementations such as high-performance relational algebra
kernels. In this paper, we present an implementation of control-flow
analysis for a significant subset of Scheme (including \texttt{set!},
\texttt{call/cc}, and primitive operations) using the Souffl\'e
Datalog engine. We present an evaluation on a worst-case term
demonstrating the polynomial complexity of our $m$-CFA and remark upon
scalability results using Souffl\'e.
\end{abstract}


\maketitle

\section{Introduction}

Static analysis is a technique to explicate properties of a program's
behavior via inspecting only the program's source code (without
executing the program)~\cite{ppa}. There exist many frameworks for
constructing program analyses, e.g., Cousot and Coust's abstract
interpretation~\cite{cousot1977abstract}. A static analysis is
\emph{sound} when it is strictly conservative in the sense that any
true program behavior is reported (at least approximately) as a result
of the analysis. Unfortunately---due to the halting problem---no
terminating static analysis may be both sound and \emph{complete} (all
reported results represent true behavior). While static analyses may
be constructed using arbitrary degrees of precision (e.g., via
instrumentation-based polyvariance~\cite{gilray2016polyvariance}) in
principle, in practice balancing precision and complexity while
retaining soundness requires significant engineering
effort~\cite{Bravenboer:09}. 

A central challenge in analyzing Scheme programs is control-flow
analysis: for each callsite, which syntactic lambdas (in which
contexts, for a context-sensitive analysis) may be invoked? This
problem is simple in procedural languages (which include only direct
control-flow), but challenging for higher-order languages, as
data-flow and control-flow must be performed simultaneously. Shivers
defined the $k$-CFA family of increasingly-precise control-flow
analyses for Scheme~\cite{shivers1991cfa}. However, as Shivers notes,
uniform $k$-CFA quickly becomes intractable for the case $k > 0$, even
for reasonably-sized programs~\cite{Olin:2004}. In fact, Van Horn and
Mairson later showed that $k$-CFA is
EXPTIME-complete~\cite{VanHorn:2008}.

Compared to Scheme, significant engineering work has been expended
into whole-program analysis for object-oriented languages,
particularly
Java~\cite{bravenboer2009doop,Scholz:2016,Kastrinis:2013,Balatsouras:2017}. This
has culminated in state-of-the-art systems such as DOOP, whose
analysis is written in Datalog for subsequent compilation to efficient
relational algebra kernels implemented via
C\texttt{++}~\cite{bravenboer2009doop}. Systems such as DOOP scale to
large (multi-thousand line) codebases even when context-sensitivity is
considered. This would appear at odds with Van Horn and Mairson's
result that $k$-CFA is EXPTIME-complete. To resolve this paradox,
Might et al. show that the degenerative closure structure of
object-oriented languages (analogous to the difference between flat
and linked closures), context-sensitive control-flow analyses of
object-oriented languages is PTIME~\cite{might2010mcfa}. The authors
also present $m$-CFA, a technique to analyze arbitrary higher-order
languages using flat closures to achieve polynomial complexity.

While the techniques for analyzing object-oriented languages using
Datalog are well understood, we are aware of no existing presentation
illustrating how Datalog may be used to implement the analysis of
Scheme. We find this particularly surprising, as we believe the
extreme efficiency of modern Datalog engines may prove a key enabling
technology to tackle the inherent complexity of higher-order
languages.

In this paper, we present a systematic approach for deriving a
Datalog-based implementation of control-flow analysis for Scheme-like
languages. Key to our approach is the formulation of the analysis via
the abstracting abstract machines (AAM) methodology of Van Horn and
Might~\cite{van2010aam}. We present how this formulation may be
translated to Datalog, overcoming several key obstacles unique to
Scheme. To evaluate our approach, we implemented an analysis of a
significant subset of Scheme including multi-argument lambdas,
conditional control-flow, builtins, and first-class continuations
(\texttt{call/cc}). In section~\ref{sec:semantics}, we present a
formalization of this language (noting how our choices anticipate a
Datalog implementation) via the AAM approach, following in
section~\ref{sec:datalog} with its corresponding Datalog
transliteration (we plan to the implementation of our analysis
open-source). We validated (manually, via inspection) the correctness
of our analysis and in section~\ref{sec:eval} present the results of a
set of experiments benchmarking our implementation of $m$-CFA for our
subset of Scheme.

\section{Background and Related Work}

In this section, we sketch several key background concepts that
underlie our abstract semantics in Section~\ref{sec:semantics} and
subsequent Datalog implementation in Section~\ref{sec:datalog}.

\subsection{Program Analysis and Abstract Interpretation}

Kildall first introduced dataflow analysis (of flowchart-style
programs in the style of Floyd~\cite{Floyd1967Flowcharts}) to
approximate static program behavior for the purpose of compile-time
optimization~\cite{Kildall:1973}. A central idea of Kildall's work was
using a lattice to impose an ordering on analysis results and ensure
termination via the finiteness of said lattice. Cousot and Cousot
later generalized Kildall's ideas to define the abstract
interpretation of a program. Abstract interpretation allows relating
an arbitrary pair of lattices, typically a concrete state space
($\Sigma$) and its abstraction ($\hat{\Sigma}$), along with a pair,
$(\alpha,\gamma)$, of (adjunctive) mappings between them for
abstraction ($\alpha$) and concretization ($\gamma$). Using this
abstraction alongside a \emph{collecting semantics} allows iterating a
program analysis to some fixed-point in an arbitrary lattice of
abstract results. Assuming this lattice of results is of finite
height, the collecting semantics (and therefore analysis) will
necessarily terminate via Tarski's fixpoint
theorem~\cite{Tarski:1955}.

We elide a complete presentation of abstract interpretation; there
exist several expository texts including those by
Min\'{e}~\cite{Mine:2017} and Nielsen and Nielsen~\cite{ppa}.

\subsection{Control Flow Analysis}

Languages such as C do not include indirect control flow. Determining
control flow (and also data flow) is simple for these languages, as
control is syntactically-apparent. In scheme, it is more difficult to
tell which values flow to a particular variable because of the
pervasive use of higher-order functions.

Consider the following Scheme code:

\begin{verbatim}
(let* ([f (foo 42)]
       [g (bar 99)]
       [h (if (= a b) (g 30) (f g))])
  (g h))
\end{verbatim}

Deciding which branch of the \texttt{if} is taken depends on (at
least) the values that flow to \texttt{a} and \texttt{b}. Similarly,
to understand data flow, we also must know control flow: knowing which
value flows to \texttt{h} requires reasoning about the \texttt{if}'s
control flow. The key is to compute both simultaneously. As the
computation continues, data flow information is fed to create a
control-flow graph (CFG) on-the-fly, and the new CFG is used to find
new data-flow. This is a central idea in the original formulation of
$k$-CFA by Shivers~\cite{shivers1991cfa}, though presentations can
also be found elsewhere~\cite{ppa,Mine:2017}.

\subsection{Abstract Abstract Machines}

Might and Van Horn presented the Abstracting Abstract Machines (AAM)
approach to abstract interpretation for functional
languages~\cite{van2010aam}. The key insight in their work is to
redirect all sources of recursion in the analysis through a store
which may be finitized by construction. Using this approach, an
abstract semantics may be derived from an abstract machine specifying
a concrete semantics. The AAM-based approach can compute any type of
CFA, and encompasses a broad array of analysis precision including,
e.g., object-sensitivity \cite{gilray2016polyvariance}. The machine
described in this paper utilizes $m$-CFA \cite{might2010mcfa}, a
variant of $k$-CFA that uses flat closures.

\subsection{Datalog}

Datalog is a bottom-up logic programming language largely based on
Horn-SAT. We refer the reader to the exposition of Ceri et al. for a
detailed description of Datalog~\cite{ceri1989datalog}. Datalog
programs consist of a set of Horn clauses of the form $P(x_0, ...)
\leftarrow Q(y_0, ...) \land ... \land S(z_0, ...)$. To evaluate these
programs, an extensional database (EDB) is provided as input
specifying a set of initial facts. A Datalog engine then runs the
rules to a fixed-point to produce an output database. The following
example computes the \texttt{cousin} relation from the EDB relations
of \texttt{parent} and \texttt{sibling}.

\begin{verbatim}
cousin(a, c) :- parent(a, p), sibling(p, q), parent(c, q).
\end{verbatim}

Relations can be recursive. Calculating ancestry is simple. The base case shows
that a parent is trivially an ancestor, but a parent of an ancestor is also an
ancestor:

\begin{verbatim}
ancestor(p, a) :- parent(p, a).
// If p already has some ancestor a,
// the parent b of a is also an ancestor of p.
ancestor(p, b) :- ancestor(p, a), parent(a, b).
\end{verbatim}

\section{Syntax and Abstract Semantics}
\label{sec:semantics}

\begin{figure}[h]
  \centering
  \begin{center}
    \textbf{Syntactic Classes}
  \end{center}
  \begin{minipage}[c]{0.45\linewidth}
    \begin{align*}
      e \in \Exp &::= \mae \\
                 & |\; \ifsyn{e}{e}{e} \;|\; \setsyn{x}{e} \\
                 & |\; \callccsyn{e} \;|\; \llet \\
                 &|\; (op \; e \; e) \;|\; (e \; e \; e \; ...) \\
      \mae \in \textsf{AExp} &::= x \;|\; \lam \;|\; b \;|\; n \\
    \end{align*}
  \end{minipage}
  \begin{minipage}[c]{0.45\linewidth}
    \begin{align*}
      b \in \mathbb{B} &\triangleq \{\truesyn , \falsesyn\} \\
      n \in \mathbb{Z} &\\
      x \in \Var &\triangleq \text{The set of identifiers} \\
      \llet \in \Let &::= \letsyn{x}{e}{e} \\
      \lam \in \Lam &::= (\lambda \; (x) \; e) \\
      \prim \in \Prim &\triangleq \text{The set of primitives}
    \end{align*}
  \end{minipage}
  \caption{Syntactic Classes for the Scheme CESK* Machine}
  \label{fig:syntax classes}
\end{figure}

Many famous papers use a significant subset of scheme, perhaps too small for
useful real-world analyses. Might et al.'s m-CFA paper uses a variant of
the CPS lambda calculus with multi-argument functions \cite{might2010mcfa}.
Shivers' original work on CFA utilized a CPS subset without conditionals or
mutation, instead relying on primitives \cite{shivers1991cfa}.
Van Horn et al.'s original Abstracting Abstract Machines adds
mutation, conditionals, and first order continuations, but does not include
multi-argument lambdas, which is a crucial component of CFA \cite{van2010aam}.
Each of these leaves out important components which are required to build
analyses on real languages.

In Figure \ref{fig:syntax classes}, we separate complex from atomic expressions.
We also include a variety of useful syntax such as mutation through
\texttt{set!} expressions, \texttt{let} bindings, and the higher order control
flow operator, \texttt{call/cc}. Conditional expressions and primitives are also
highly important for writing meaningful examples. We support binary primitives,
which are needed to analyze real scheme programs. Primitive \texttt{op}s include
recursive data (i.e. lists), mathematical, and logical operations.

Our subset of scheme is capable of supporting many real world programs. Analyses
will be more understandable and easier to implement due to these features.
This syntax also clarifies contexts in a CFA and how they grow and shrink. With
\texttt{let} and multi-argument lambdas, we can place multiple bindings in a
single context. This will avoid states ascending to top, by decreasing the
amount of contexts.

\begin{figure}[h]
  \centering
  \begin{center}
    \textbf{Semantic Classes}
  \end{center}
  \begin{minipage}[t]{0.45\linewidth}
    \begin{align*}
      \sstate \in \State &\triangleq E\langle \Eval \rangle
                           + A\langle \Apply \rangle \\
      \Eval &\triangleq \Exp \times \Context \\
                                    & \times \Store \times \KAddr \\
      \Apply &\triangleq \Val \times \Store \times \KAddr \\
      \store \in \Store &\triangleq \VStore \times \KStore \\
      \store_v \in \VStore &\triangleq \VAddr
                             \rightharpoonup \mathcal{P}(\Val) \\
      \store_k \in \KStore &\triangleq \KAddr
                             \rightharpoonup \mathcal{P}(\Kont) \\
      \context \in \Context &\triangleq \Exp^m \\
      \vaddr \in \VAddr &\triangleq \Var \times \Context \\
      \kaddr \in \KAddr &\triangleq \Exp \times \Context \\
    \end{align*}
  \end{minipage}
  \begin{minipage}[t]{0.45\linewidth}
    \begin{align*}
      \val \in \Val &\triangleq \mathbb{Z} + \mathbb{B} + \PrimPoint
                      + \Closure + \KAddr \\
      \primPoint \in \PrimPoint &\triangleq \Prim \times \Val \times \Val \\
      \closure \in \Closure &\triangleq \Lam \times \Context \\
      \kont \in \Kont &::= \mtk \;|\; \ifk{e}{e}{\context}{\kaddr} \\
                    &  |\; \setk{\vaddr}{\kaddr}
                      \;|\; \callcck{\context}{\kaddr} \\
                    & |\; \letk{\vaddr}{e}{\context}{\kaddr}
                      \;|\; \fnk{v}{n}{\context}{\kaddr} \\
                    & |\; \primsk{\prim}{e}{\context}{\kaddr}
                      \;|\; \primfk{\prim}{\val}{\kaddr} \\
                    &  |\; \argk{e\;...}{\context}{\context}{\kaddr}
    \end{align*}
  \end{minipage}
	\caption{Semantic Classes for the Scheme m-CFA CESK* Machine}
  \label{fig:semantic classes}
\end{figure}


Figure \ref{fig:semantic classes} presents a CESK* machine utilizing $m$-CFA for
value-analysis, \cite{van2010aam} \cite{might2010mcfa}. The states are
partitioned into \textit{evaluation} and \textit{application} states, and the
store is partitioned into a value-store and a continuation-store. There are a
variety of continuations which give meaning to the program.

The state is partitioned into 2 types of sub-state. The control of an
\textit{Eval} state is syntax. When transitioning from an \textit{Eval} state,
the goal is to produce a value. For example, when an \texttt{if} expression is
encountered, the resulting state is another \textit{Eval} state with the guard
expression as the control. The second type of state is an \textit{Apply}, where
the control is a value. These states will apply the control depending on the
continuation. When an \texttt{if} expression's guard reaches an \textit{Apply}
state, a branch is selected based on the inspected value.

With abstracted abstract machines, both values and continuations are identified
through the store with an address. However, they require different address
types. Value addresses are determined by the polyvariance and analyses type we
are conducting \cite{gilray2016polyvariance}. Continuations, however, are
allocated in the Pushdown-For-Free style, which has its own allocator
\cite{gilray2016p4f}. These stores are combined in the state, but are accessed
separately and combined when clear in the semantics for brevity.

Environments in this machine are not a mapping, as shown in many AAM based
approaches. Instead, environments are simply a function of context, and stand-in
for time-stamps a la $k$-CFA \cite{might2010mcfa}. Here, closures utilize
flat-contexts, instead of the usual linked model. Instead of store-fetches
being $\store(\context(x))$, they are now $\store(x, \context)$. In $m$-CFA,
the current context is based on the top $m$ stack frames, as opposed to the
\textit{latest} $k$ call-sites.

There are a variety of continuation types to enumerate the various semantic
features of the Scheme subset. When a \texttt{set!} expression is evaluated, a
$\widehat{\textbf{setk}}$ continuation is added to the store to identify what to
accomplish after the inner expression is fully evaluated.

\begin{figure}[h]
  \begin{center}
    \textbf{Evaluation Rules}
  \end{center}
  \begin{minipage}[t]{0.49\linewidth}
    \begin{center}
      $\E{\ifsyn{e_g}{e_t}{e_f}}{\context}{\store}{\kaddr}$
      $\leadsto \E{e_g}{\context}{\store'}{\kaddr'}$
    \end{center}
    \vspace{-2mm}
    \begin{align*}
      \text{where } \tag{\textbf{E-If}}
      \kaddr' &\triangleq \allock(\sstate, e_c , \context) \\
      \kont &\triangleq \ifk{e_t}{e_f}{\context}{\kaddr} \\
      \kstore' &\triangleq \kstore \sqcup [\kaddr' \mapsto \kont]
    \end{align*}
    \vspace{0mm}
    \begin{center}
      $\E{(\prim \; e_0 \; e_1)}{\context}{\store}{\kaddr}$
      $\leadsto \E{e_0}{\context}{\store'}{\kaddr'}$
    \end{center}
    \vspace{-2mm}
    \begin{align*}
      \text{where } \tag{\textbf{E-Prim}}
      \kaddr' &\triangleq \allock(\sstate, e_i, \context) \\
      \kont &\triangleq \primsk{\prim}{e_1}{\context}{\kaddr} \\
      \kstore' &\triangleq \kstore \sqcup [\kaddr' \mapsto \kont]
    \end{align*}
    \vspace{0mm}
    \begin{center}
      $\E{\llet}{\context}{\store}{\kaddr}$
      $\leadsto \E{e_i}{\context}{\store'}{\kaddr'}$
    \end{center}
    \vspace{-2mm}
    \begin{align*}
      \text{where } \tag{\textbf{E-Let}}
      \context' &\triangleq \new(\sstate) \\
      \llet &= (\texttt{let} \; ((x_0 \; e_0) \; (x_s \; e_s) \; ...) \; e_b) \\
      (x_i, e_i) &\in ([x_0 : x_s] , [e_0 : e_s]) \\
      \vaddr &\triangleq \allocv(x_i , \sstate) \\
      \kaddr' &\triangleq \allock(\sstate, e_i, \context) \\
      \kont &\triangleq \letk{e_b}{\vaddr}{\context'}{\kaddr} \\
      \kstore' &\triangleq \kstore \sqcup [\kaddr' \mapsto \kont]
    \end{align*}
  \end{minipage}
  \begin{minipage}[t]{0.49\linewidth}
    \begin{center}
      $\E{\callccsyn{e}}{\context}{\store}{\kaddr}$
      $\leadsto \E{e}{\context}{\store'}{\kaddr'}$
    \end{center}
    \vspace{-2mm}
    \begin{align*}
      \text{where } \tag{\textbf{E-C/cc}}
      \context' &\triangleq \new(\sstate) \\
      \kaddr' &\triangleq \allock(\sstate, e , \context) \\
      \kont &\triangleq \callcck{\context'}{\kaddr} \\
      \kstore' &\triangleq \kstore \sqcup [\kaddr' \mapsto \kont]
    \end{align*}
    \vspace{0mm}
    \begin{center}
      $\E{\setsyn{x}{e}}{\context}{\store}{\kaddr}$
      $\leadsto \E{e}{\context}{\store'}{\kaddr'}$
    \end{center}
    \vspace{-2mm}
    \begin{align*}
      \text{where } \tag{\textbf{E-Set!}}
      \kaddr' &\triangleq \allock(\sstate, \setsyn{x}{e}, \context)  \\
      \vaddr &\triangleq \allocv(\setsyn{x}{e}, \sstate) \\
      \kont &\triangleq \setk{\vaddr}{\kaddr} \\
      \kstore' &\triangleq \kstore \sqcup [\kaddr' \mapsto \kont]
    \end{align*}
    \vspace{0mm}
    \begin{center}
      $\E{(e_f \; e_0 \; e_s \; ...)}{\context}{\store}{\kaddr}$
      $\leadsto \E{e_f}{\context}{\store'}{\kaddr'}$
    \end{center}
    \vspace{-2mm}
    \begin{align*}
      \text{where } \tag{\textbf{E-Call}}
      \context' &\triangleq \new(\sstate) \\
      \kaddr' &\triangleq \allock(\sstate,
                (e_f \; e_0 \; e_s \; ...), \context', \kaddr ) \\
      \kont &\triangleq \argk{[e_0 : e_s]}{\context}{\context'}{\kont} \\
      \kstore' &\triangleq \kstore \sqcup [\kaddr' \mapsto \kont]
    \end{align*}
  \end{minipage}
	\caption{Rules to evaluate syntax into smaller expressions}
  \label{fig:Eval Rules}
\end{figure}

The \textit{Eval} rules in Figure \ref{fig:Eval Rules} govern how to break down
complex expressions to form an atomic expression. Some rules are
straightforward, such as \textbf{E-If}. After encountering an \texttt{if}
expression, the guard is evaluated. A continuation is created to keep track of
what to do when we know the value of the guard. Generally, an evaluation rule
should focus on evaluating syntax, and not creating semantic objects such as
contexts.

Although, other transition rules are trickier. The \textbf{E-Let} transitions to
multiple states, one for each binding. For brevity, at least one binding is
required, as a 0-binding \texttt{let} would have to transition directly to the
body with a new context. Let is also complex since it creates the resulting
context

The \textbf{E-Prim} and \textbf{E-Call} rules are
multi-stage rules. They each have multiple expressions that must be evaluated in
a specific order. In a function call, the function is evaluated before the
arguments, so there must be a special continuation frame for each juncture in
the evaluation.

\begin{figure}[h]
	\begin{center}
    \textbf{Atomic Evaluation}
  \end{center}
  \begin{minipage}[c]{0.45\linewidth}
    \begin{center}
      $\E{\mae}{\_}{\store}{\kaddr}$
      $\leadsto \A{\val}{\store}{\kaddr}$
    \end{center}
    \vspace{-2mm}
    \begin{align*}
      \text{where } \tag{\textbf{E-AE}}
      \val &\triangleq \aeval(\sstate) \\
    \end{align*}
  \end{minipage}
  \begin{minipage}[c]{0.45\linewidth}
    \begin{align*}
      \aeval :: \Eval &\rightarrow \val\\
      \aeval(b, \_, \_) \triangleq b \;&\; \aeval(n, \_, \_) \triangleq n\\
      \aeval(x, \context, \_) &\overset{\triangle}{\ni} \vstore(x, \context) \\
      \aeval(lam, \context, \_) &\triangleq (lam, \context) \\
    \end{align*}
  \end{minipage}
	\caption{Atomic Evaluation, which converts syntax into a value.}
  \label{fig:Atomic Eval}
\end{figure}

Atomic evaluation is defined in Figure \ref{fig:Atomic Eval}. The function
$\aeval$ produces values from atomic expressions. The rule \textbf{E-AE} will
transition from an \textit{Eval} state into an \textit{Apply} state using
$\aeval$.

\begin{figure}[h]
  \begin{center}
    \textbf{Application Rules}
  \end{center}
  \begin{minipage}[t]{0.49\linewidth}
    \begin{center}
      $\A{\val}{\store}{\kaddr}$
      $\leadsto \E{e_t}{\context}{\store}{\kaddr'}$
    \end{center}
    \vspace{-2mm}
    \begin{align*}
      \text{where } \tag{\textbf{A-IfT}}
      \kstore(\kaddr)  &\ni \ifk{e_t}{\_}{\context}{\kaddr'} \\
      \val &\neq \falsesyn
    \end{align*}
    \vspace{0mm}
    \begin{center}
      $\A{\val}{\store}{\kaddr}$
      $\leadsto \E{e_f}{\context}{\store}{\kaddr'}$
    \end{center}
    \vspace{-2mm}
    \begin{align*}
      \text{where } \tag{\textbf{A-IfF}}
      \kstore(\kaddr) &\ni \ifk{\_}{e_f}{\context}{\kaddr'} \\
      \val &= \falsesyn
    \end{align*}
    \vspace{0mm}
    \begin{center}
      $\A{\val}{\store}{\kaddr}$
      $\leadsto \E{e_b}{\context}{\store'}{\kaddr'}$
    \end{center}
    \vspace{-1mm}
    \begin{align*}
      \text{where } \tag{\textbf{A-Let}}
      \kstore(\kaddr) &\ni \letk{e_b}{\vaddr}{\context}{\kaddr'} \\
      \vstore' &\triangleq \vstore \sqcup [\vaddr \mapsto \val]
    \end{align*}
    \vspace{0mm}
    \begin{center}
      $\A{\val}{\store}{\kaddr}$
      $\leadsto \A{-42}{\store}{\kaddr'}$
    \end{center}
    \vspace{-2mm}
    \begin{align*}
      \text{where } \tag{\textbf{A-Set!}}
      \kstore(\kaddr) &\ni \setk{\vaddr}{\kaddr'} \\
      \vstore' &\triangleq \vstore \sqcup [\vaddr \mapsto \val]
    \end{align*}
    \vspace{0mm}
    \begin{center}
      $\A{\val}{\store}{\kaddr}$
      $\leadsto \E{e}{\context}{\store'}{\kaddr''}$
    \end{center}
    \vspace{-1mm}
    \begin{align*}
      \text{where } \tag{\textbf{A-Prim1}}
      \kstore(\kaddr) &\ni \primsk{\prim}{e}{\context}{\kaddr'} \\
      \kaddr'' &\triangleq \allock(\sstate, e, \context) \\
      \kont &\triangleq \primfk{\prim}{\val}{\kaddr'} \\
      \kstore' &\triangleq \kstore \sqcup [\kaddr'' \mapsto \kont]
    \end{align*}
    \vspace{0mm}
    \begin{center}
      $\A{\val'}{\store}{\kaddr}$
      $\leadsto \A{\textbf{primVal}(\prim, \val, \val')}{\store}{\kaddr'}$
    \end{center}
    \vspace{-1mm}
    \begin{align*}
      \text{where } \tag{\textbf{A-Prim2}}
      \kstore(\kaddr) &\ni \primfk{\prim}{v}{\kaddr'} \\
    \end{align*}
  \end{minipage}
  \begin{minipage}[t]{0.49\linewidth}
    \begin{center}
      $\A{\val}{\store}{\kaddr}$
      $\leadsto \E{e_b}{\context}{\store''}{\kaddr'}$
    \end{center}
    \vspace{-2mm}
    \begin{align*}
      \text{where } \tag{\textbf{A-C/cc}}
      \kstore(\kaddr) &\ni \callcck{\context}{\kaddr'} \\
      \val &= ((\lambda \; (x) \; e_b), \context_{\closure}) \\
      \vaddr &\triangleq \allocv (x, \sstate) \\
      \vstore' &\triangleq \copyctx(\context_{\closure}, \context) \\
      \vstore'' &\triangleq \vstore \sqcup [\vaddr \mapsto \kaddr]
    \end{align*}
    \vspace{0mm}
    \begin{center}
      $\A{\kaddr}{\store}{\kaddr'}$
      $\leadsto \A{\kaddr'}{\store}{\kaddr}$
    \end{center}
    \vspace{-2mm}
    \begin{align*}
      \text{where } \tag{\textbf{A-C/ccKont}}
      \kstore(\kaddr') &\ni \callcck{\_}{\_}
    \end{align*}
    \vspace{0mm}
    \begin{center}
      $\A{\val}{\store}{\kaddr}$
      $\leadsto \E{e_i}{\context}{\store'}{\kaddr'}$
    \end{center}
    \vspace{-2mm}
    \begin{align*}
      \text{where } \tag{\textbf{A-Ar}}
      \kstore(\kaddr) &\ni \argk{e_s}{\context}{\context'}{\kaddr'} \\
      e_i &\in e_s \\
      \kaddr' &\triangleq \allock(\sstate, e_i, \context) \\
      \kont &\triangleq \fnk{\val}{i}{\context'}{\kaddr'} \\
      \kstore' &\triangleq \kstore \sqcup [\vaddr \mapsto \kont]
    \end{align*}
    \vspace{0mm}
    \begin{center}
      $\A{\val}{\store}{\kaddr}$
      $\leadsto \E{e_b}{\context}{\store''}{\kaddr'}$
    \end{center}
    \vspace{-2mm}
    \begin{align*}
      \text{where } \tag{\textbf{A-Call}}
      \kstore(\kaddr) &\ni \fnk{\closure}{n}{\context}{\kaddr'} \\
      \closure &= ((\lambda \; (x_s ...) \; e_b) , \context_{\closure}) \\
      \vaddr &\triangleq \allocv(x_n, \sstate) \\
      \vstore' &\triangleq \copyctx(\context_{\closure}, \context) \\
      \vstore'' &\triangleq \vstore \sqcup [\vaddr \mapsto \val]
    \end{align*}
    \vspace{0mm}
    \begin{center}
      $\A{\val}{\store}{\kaddr}$
      $\leadsto \A{\val}{\store}{\kaddr'}$
    \end{center}
    \vspace{-2mm}
    \begin{align*}
      \text{where } \tag{\textbf{A-CallKont}}
      \kstore(\kaddr) &\ni \fnk{\kaddr'}{\_}{\_}{\_} \\
    \end{align*}
  \end{minipage}
	\caption{Value Application Rules}
  \label{fig:Apply Rules}
\end{figure}

Apply rules, as shown in Figure \ref{fig:Apply Rules} transition based on the
value and the current continuation. These rules primarily govern contexts. Here,
the context may be extended, bindings added to it, or it may be returned to a
previous version.

Apply rules may be triggered at the same time. For example, if an address
contains multiple value, then both \textbf{A-If} rules must be transitioned
from. In Figure \ref{fig:If Conflate}, both branches must be taken to
soundly approximate this program when using a low sensitivity analysis
such as 0-CFA. In the example, the inner let calls the same function twice,
which, in the same context, binds the address $x$ to both \truesyn
\hspace{0.25mm} and \falsesyn. Then, when $a$ is referenced, it contains both
possibilities, so both branches will be taken.

\begin{figure}
  \centering
  \begin{Verbatim}
                            (let ([f (lambda (x) x)])
                              (let ([a (f #t)] [b (f #f)])
                                (if a 4 5)))
  \end{Verbatim}
  \caption{An example where both branches of a condition must be taken.}
  \label{fig:If Conflate}
\end{figure}

The \textbf{A-Let} rule will be ran once for every binding in a \texttt{let}
expression. Because the addresses were calculated in the \textbf{E-Let} rule,
not much work needs to be done in this rule. However, because every
$\widehat{\textbf{let}}$ continuation results in the same state, only one output
state is generated for the body of the expression.

Creating new contexts is only done when variables are bound. Creating scope like
this is the trigger, and not simply evaluating an expression. Therefore, we only
need to create a context in the case of \textbf{E-Let}, \textbf{E-Call/cc}, and
\textbf{E-Call}. We only set the context in the \textit{Apply} state, however,
when we are evaluating the inner expression with the new bindings.



\begin{figure}[h]
	\begin{center}
    \textbf{Helper Functions}
  \end{center}
  \begin{minipage}[c]{0.45\linewidth}
    \begin{center}
      $\inj :: \Exp \rightarrow \State$ \\
      $\inj(e) \triangleq \E{e}{\epsilon}{(\bot , \kstore)}{(e, \epsilon)}$ \\
      $\kstore \triangleq \bot \sqcup [(e, \epsilon) \mapsto \mtk]$ \\
      $ $ \\
      $\new :: \Eval \rightarrow \Context$ \\
      $\new(e, \context, \_) \triangleq \lfloor e:context \rfloor_m$
    \end{center}
  \end{minipage}
  \begin{minipage}[c]{0.45\linewidth}
    \begin{center}
      $\allocv :: \Var \times \State \rightharpoonup \VAddr$ \\
      $\allocv(x, \E{\_}{\context}{\_}) \triangleq (x, \context)$ \\
      $\allocv(x, \A{(\_, \context)}{\_}) \triangleq (x, \context)$ \\
      $ $ \\
      $\allock :: \State \times \Exp \times \Context  \rightharpoonup \KAddr$ \\
      $\allock(\_, e, \context) \triangleq (e, \context)$ \\
    \end{center}
  \end{minipage}
	\caption{Auxiliary functions}
  \label{fig:Aux Functions}
\end{figure}

\begin{figure}[h]
	\begin{center}
    \textbf{Transition Relation and Global Store}
  \end{center}
  \begin{minipage}[c]{0.45\linewidth}
    \begin{align*}
      (\leadsto) :: \State &\rightharpoonup \mathcal{P}(\State) \\
      (\leadsto_{\StateSpace}) :: \StateSpace &\rightharpoonup \mathcal{P}(\StateSpace) \\
      (\reachableConfig, \store) &\leadsto_{\StateSpace} (\reachableConfig', \store')\\
    \end{align*}
    \vspace{-11mm}
    \begin{align*}
      \text{where }
      \hat{s} &\triangleq \{\sstate \;\vert\; (r, \store) \leadsto \sstate \}
                \cup \{\inj(e_o)\ \\
      \reachableConfig' &\triangleq \{r \;\vert\; (r, \_) \in \hat{s} \} \\
      \store' &\triangleq \bigsqcup_{(\_, \store'') \in \hat{s}} \store''
    \end{align*}
  \end{minipage}
  \begin{minipage}[c]{0.45\linewidth}
    \begin{align*}
      \stateSpace \in \StateSpace &\triangleq \ReachableConfig \times \Store \\
      \reachableConfig \in \ReachableConfig &\triangleq \mathcal{P}(\Config) \\
      \config \in \Config &\triangleq
                            CE\langle \Exp \times \Context \times \KAddr\rangle \\
                                  & + CA\langle \Val \times \KAddr \rangle
    \end{align*}
  \end{minipage}
	\caption{Transition Relation and Global Store}
  \label{fig:Global Store}
\end{figure}

The value and continuation stores are necessarily responsible for an exponential
growth of states in the state space. One method to maintain polynomial
worst-case complexity when evaluating closures,
is to separate the stores from the states, and to globalize the stores.
After each state transition, the stores are combined and
used for the next transitions. Figure \ref{fig:Global Store} gives a collecting
semantics which transforms a standard machine with an in-line store into a
global store. The semantics utilize the injection function $\inj$, along with
the inline-store transition function $\leadsto$.

\section{Datalog Implementation}
\label{sec:datalog}

Souffl\'e is used as the Datalog engine for its state-of-the-art runtime
performance, parallelism, and language features. Algebraic Data Types (ADT) are
utilized for contexts, value and continuation types to maintain brevity. This
implementation can be replicated without ADTs but it will be much more verbose.

Implementing the above operational semantics as a Datalog program is generally a
straightforward process. Many rules can be converted in a near 1:1 fashion.
However, there are some unique relations that differ from the operational
semantics. Still, the rules of the machine map closely to the implementation
in Datalog. Figure \ref{fig:EIf} shows the similarity between the two.

\begin{figure}
\begin{minipage}[c]{0.45\linewidth}
  \begin{verbatim}
state_e(eguard, ctx, ak),
stored_kont(ak, kont),
flow_ee(e, eguard) :-
    state_e(e, ctx, ak),
    if(e, eguard, et, ef),
    ak = $KAddress(eguard, ctx),
    kont = $If(et, ef, ctx, ak).
  \end{verbatim}
\end{minipage}
\begin{minipage}[c]{0.49\linewidth}
  \begin{center}
    $\E{\ifsyn{e_g}{e_t}{e_f}}{\context}{\store}{\kaddr}$
    $\leadsto \E{e_g}{\context}{\store'}{\kaddr'}$
  \end{center}
  \vspace{-2mm}
  \begin{align*}
    \text{where } \tag{\textbf{E-If}}
    \kaddr' &\triangleq \allock(\sstate, e_c , \context) \\
    \kont &\triangleq \ifk{e_t}{e_f}{\context}{\kaddr} \\
    \kstore' &\triangleq \kstore \sqcup [\kaddr' \mapsto \kont]
  \end{align*}
\end{minipage}
\caption{If expression evaluation in Datalog and in the operational semantics.}
\label{fig:EIf}
\end{figure}

Datalog programs being split into rules maps well onto
operational semantics. There is a correspondence between the operational
semantics and the Datalog implementation. Datalog places the head of the Horn
clause before the body, so the output state is on top, along with the
continuation being added to the store. The body of the Datalog rule acts as the
input. As rules are computed, new facts are generated by the heads of the
clauses. As a result, more rules can be executed, extending the set of total
facts.

Figure \ref{fig:copy and peek} shows two interesting relations. These highlight
a key difference between the operational semantics and the Datalog
implementation. On the left is the equivalent to $\new$: \texttt{peek\_ctx}. On
the right is the implementation of $\copyctx$.
In Datalog, we compute a helper relation, \texttt{peek\_ctx}, which computes
a context as needed. The \texttt{copy\_ctx} relation will signal a copy
operation, from a source to a destination context. When a \texttt{copy\_ctx}
fact is added, it does not immediately do the copy, but it will trigger an
inference rule for the \texttt{stored\_val} relation.

\begin{figure}
  \begin{minipage}[c]{0.45\linewidth}
\begin{verbatim}
peek_ctx(e, old_ctx, new_ctx) :-
    state_e(e, old_ctx, _),
    ( callcc(e, _) ; call(e, _, _)
    ; let(e, _, _); lambda(e,_,_)),
    old_ctx = $Context(ctx1,ctx0),
    new_ctx = $Context(e,ctx1).
\end{verbatim}
  \end{minipage}
  \begin{minipage}[c]{0.45\linewidth}
\begin{verbatim}
stored_val(av, v) :-
    copy_ctx(from, to, e),
    freevar(fv, e),
    stored_val(av, v),
    av = $VAddress(fv, to).
\end{verbatim}
  \end{minipage}
  \caption{The new and copy function analogues in the Datalog implementation.}
  \label{fig:copy and peek}
\end{figure}

See Appendix A for the full Datalog implementation.

\section{Evaluation}
\label{sec:eval}

In this section we present an evaluation of our implementation of
$m$-CFA in Souffl\'e. To gain confidence in our implementation's
correctness, we performed manual validation on a set of testcases. To
measure scalability and complexity in practice, we construct a family
of terms that exhibit worst-case (polynomial) behavior. We used large
terms in this family and ran experiments on a 28-core Linux server
with 78GB of RAM.

\subsection{Constructing worst-case terms for $m$-CFA}

In order to both evaluate the performance and verify the correctness
of our implementation, we construct a family of terms that incur
maximum-possible work in practice. Van Horn’s dissertation
details a construction of terms whose analysis require exponential
work for $k$-CFA~\cite{VanHorn:2008}. Figure~\ref{fig:dvh-exptime}
shows an example for 1-CFA. The ultimate issue is that \texttt{\#t} and
\texttt{\#f} will be conflated through the call to \texttt{f}, producing
two runtime closures but four abstract closures; more bindings, or
arguments to $w$, may be added to add further complexity.

We use Van Horn's technique to generate high-complexity terms for our
$m$-CFA implementation. The key trick to fool [k=1]-CFA from Van Horn's
example is the application of the identity function inside the term. This
identity function acts as \emph{padding}, as [k=1]-CFA traces only the
\emph{most recent} call site. During concrete execution, the
intermediate call to the identity function sits (on the stack) in
front of the (separate) calls to \texttt{f}. In [k=1]-CFA, only the most
recent callsite is remembered and thus \texttt{\#t} and \texttt{\#f}
will flow to same address. However we cannot directly translatiterate this
example into $m$-CFA as we will not see the expected storage blowup.
In $m$-CFA, the context (the contour in $k$-CFA) only grows in the state when
values are bound to a variable. If padding is in function postion when
\texttt{$\lambda$ (w) (w x)} is evaluated, the context has not been extended
yet. Using Van Horn's example with $m$-CFA, the padding will be bypassed.
In $m$-CFA, the padding should be moved into argument position and $\eta$-expand
the precision losing term. This is shown in Figure \ref{fig:mcfa-lost}.
This modification forces conflation of the values after the padding label is
appended into the context.


\begin{figure}
\begin{verbatim}
((lambda (f) (let ((m (f #t))
                   (n (f #f)))
          m))
 (lambda (z)
   ((lambda (x) x)
    (lambda (w) (w z z)))))
\end{verbatim}
    \caption{Example (from Van Horn) showing exponential behavior for k-CFA}
    \label{fig:dvh-exptime}
  \end{figure}

  \begin{figure}
\begin{verbatim}
((lambda (f) (let ((mm (f M)
                    ...
                   (m1 (f 1))
                   (n0 (f 0)))
          m))
 (lambda (z)
   ((lambda (x) (+ z (+ z ...)))
    (lambda (x) x))))
\end{verbatim}
    \caption{An example term from our experiments.}
    \label{fig:mcfa-lost}
\end{figure}

\begin{table}
  \caption{Running time and memory usage of $m$-CFA in Datalog. Term
    size N/K means N calls to \texttt{f} and K invocations of
    \texttt{+}.}
  \label{tab:mcfa-time}
  \begin{tabular}{cccccccc}
      \toprule
      & &  \multicolumn{2}{c}{$0\;padding $} & \multicolumn{2}{c}{$1\;padding $} & \multicolumn{2}{c}{$2\;padding $} \\
      Term size &  Polyvariance ($m$) & Time & Memory & Time & Memory & Time  &  Memeory \\
      \midrule
               &  0 &   00:09:57   &   1.27GB   & 00:09:46    &  1.86GB      & 09:48.04   & 1.27GB  \\
      32/4     &  1 &   < 1 sec    &   12.1MB   & 00:22:49    &  1.28GB      & 13:26.88   & 1.27GB  \\
               &  2 &   < 1 sec    &   12.29MB  & < 1 sec     &  12.5MB      & 19:09.89   & 1.27GB \\
      \midrule
               &  0 &  01:08:58    &   3.55GB   & 01:09:36    &  3.55GB      & 01:02:51   & 3.56GB  \\
      86/3     &  1 &  < 1 sec     &   6.31MB   & 01:34:37    &  3.55GB      & 01:32:10   & 3.56GB \\
               &  2 &  < 1 sec     &   6.31MB   & < 1 sec     &  6.32MB      & 02:13:10   & 3.56GB \\
    \bottomrule
  \end{tabular}
\end{table}

\subsection{Results}

We used our worst-case term construction to perform a variety of runs
using Souffl\'e. Table \ref{tab:mcfa-time} shows the results of our system on
a set of two terms: one with 32 \texttt{let} bindings and one with 86
\texttt{let} bindings. We measure 0, 1, and 2-CFA using a variety of
paddings (0, 1, and 2). By construction, we expect terms to explode
when the amount of padding is too low as the analysis begins to
conflate polynomially-greater callsites. For example, in 32/4, we see
0-CFA explode while 1 and 2-CFA are very fast. This is because both 1
and 2-CFA are fully-precise for the term. As expected, analyses of
higher-complexity have longer runtimes. For example, looking at the
timings for 2 padding, we can see that 2-CFA is roughly twice as slow
as 0-CFA.

\begin{table}
  \caption{ Parallel Performance of m-CFA souffle implementation. 32
    different clauses in let with 2 padding }
  \label{tab:parallel}
  \begin{tabular}{cccccccccc}
    \toprule
     &  \multicolumn{2}{c}{1 core} & \multicolumn{2}{c}{2 threads} & \multicolumn{2}{c}{4 threads} & \multicolumn{2}{c}{8 threads} \\
    $m$ & time & memory & time & memory & time  &  memeory  & time & memory\\
    \midrule
    0 &   00:09:48    &   1.27GB   &   00:13:05  &   1.43GB    & 00:18:39  & 1.55GB & 00:22:38 & 1.61GB \\
    1 &   00:13:26    &   1.27GB   &   00:14:36  &   1.46GB    & 00:20:36  & 1.55GB & 00:22:54 & 1.62GB \\
    2 &   00:19:09    &   1.27GB   &   00:25:14  &   1.46GB    & 00:36:11  & 1.54GB & 00:46:57 & 1.60GB \\

    \bottomrule
  \end{tabular}
\end{table}

We measured parallel performance using a variety of thread
counts. Souffl\'e was built to support scalability and multithreading
to speed up parallel execution on a single node. In our analysis, we
believed there would be many potentially-parallelizable states. For
exapmle, we evaluate different \texttt{let} clauses
non-deterministically. We used Souffl\'e version 2.02, compiling
from source with OpenMP support enabled. Our results were all compiled
to \texttt{C++} via Souffle\'e's \texttt{-c} flag. We verified CPU
utilization (via \texttt{htop}) to ensure multithreading was
enabled. However, as detailed in Table 2, our results demonstrated
anti-scalability: instead of making the program run faster, the more
cores used, the slower execution we observed, while also adding more
memory overhead. GitHub issues from the Souffl\'e authors point to
several potential reasons for poor scalability. For example, our
implementation uses a large number of rules, and Souffl\'e does not
parallelize across rules.

\section{Conclusion}
\label{sec:conclusion}

Datalog-based implementation of static analysis tools has enabled new
frontiers in the scalability of analyses to object-oriented
languages. However, we do not know of any presentations that extend
these ideas to Scheme-like languages. This paper presented the key
ideas necessary to implement analyses of Scheme-like languages using
Datalog. We structure our analysis using the AAM-based approach to
facilitate translation to Datalog's deductive rules, using the $m$-CFA
allocation strategy of Might et al. to mirror the flat closure
structure naturally enabled by Datalog~\cite{might2010mcfa}. To our
knowledge, this is the first presentation of a Datalog-based analysis
for Scheme-like languages.


\bibliographystyle{ACM-Reference-Format}
\bibliography{ref}

\appendix

\section{Full Souffl\'e Implementation}

\begin{Verbatim}[fontsize=\tiny]
.type id <: symbol

.type context = Context{ctx0:id,ctx1:id}

.type value = Number { n : number }
                | Bool { b : symbol }
                | Kont { k : address_k }
                | Closure { e: id, ctx: context }
                | PrimVal {op: id , v1: value , v2: value}

.type kont = MT {}
           | Arg {args: id, ctx: context, ectx: context, next_ak: address_k}
           | Fn {fn: value, pos: number, ctx: context, next_ak: address_k}
           | Set {loc: address_v, next_ak: address_k}
           | If {true_branch: id, false_branch: id,
           		     ctx: context, next_ak: address_k}
           | Callcc {ectx: context, next_ak: address_k}
           | Let {av: address_v, ebody: id, ctx: context, next_ak: address_k}
           | Prim1 {op: id, e2: id, ctx: context, next_ak: address_k}
           | Prim2 {op: id, v1: value, next_ak: address_k}

.type address_k = KAddress{e: id, ctx: context}
.type address_v = VAddress{x: symbol, ctx: context}

.decl state_e(e: id, ctx: context, ak: address_k)
.output state_e

.decl state_a(v: value, ak: address_k)
.output state_a

.decl stored_val(av: address_v, v: value)
.output stored_val

.decl stored_kont(ak: address_k, k: kont)
.output stored_kont

.decl peek_ctx(e: id, ctx_old: context, ctx_new: context)

.decl top_exp(Id: id)
.input top_exp
.decl lambda(Id: id, Vars: id, BodyId: id)
.input lambda
.decl lambda_arg_list(Id: id, Pos: number, X: symbol)
.input lambda_arg_list
.decl prim(Id: id, OpName: symbol)
.input prim
.decl prim_call(Id: id, PrimId: id, Args: id)
.input prim_call
.decl call(Id: id, FuncId: id, Args: id)
.input call
.decl call_arg_list(Id: id, Pos: number, X: id)
.input call_arg_list
.decl var(Id: id, MetaName: symbol)
.input var
.decl num(Id: id, v: number)
.input num
.decl bool(Id: id, v: symbol)
.input bool
.decl quotation(Id: id, Expr: id)
.input quotation
.decl value_form(Id: id)
value_form(id) :-
    (num(id, _); var(id, _); lambda(id, _, _); quotation(id, _); bool(id, _)).
.decl if(Id:id, GuardId: id, TrueId: id, False: id)
.input if
.decl setb(Id: id, Var: symbol, ExprId: id)
.input setb
.decl callcc(Id: id, ExprId: id)
.input callcc
.decl let(Id: id, BindId: id, BodyId: id)
.input let
.decl let_list(Id: id, X: symbol, EId: id)
.input let_list

// context changes start
state_e(e, $Context("",""), $KAddress(e, $Context("",""))),
peek_ctx(e, $Context("",""), $Context(e,"")),
stored_kont($KAddress(e, $Context("","")), $MT) :-
    top_exp(e).

peek_ctx(e, $Context(ctx1,ctx0), $Context(e,ctx1)) :-
    state_e(e, $Context(ctx1,ctx0), _),
    (callcc(e, _) ; call(e, _, _)
    ; let(e, _, _); lambda(e,_,_)).
// context end

.decl freevar(x:symbol, e: id)
freevar(x, e) :- var(e, x).
freevar(x, e) :-
    lambda(e, vars, body),
    freevar(x, body),
    lambda_arg_list(vars, _, v),
    x != v.
freevar(x, e) :-
    call(e, func, args),
    (freevar(x, func); freevar(x, args)).
freevar(x, e) :-
    prim_call(e, _, args),
    freevar(x, args).
freevar(x, e) :-
    call_arg_list(e, pos, arg),
    freevar(x, arg).
freevar(x, e) :-
    if(e, eguard, et, ef),
    (freevar(x, eguard); freevar(x, et); freevar(x, ef)).
freevar(y, e) :-
    setb(e, _, ev),
    freevar(y, ev).
freevar(x, e) :-
    callcc(e, ev),
    freevar(x, ev).
freevar(x, e) :-
    let(e, binds, body),
    (freevar(x, binds); freevar(x, body)).
freevar(x, e) :-
    let_list(e, a, bind),
    freevar(x, bind),
    x != a.

.decl flow_ee(e1: id, e2: id)
.output flow_ee
.decl flow_ea(e1: id, a2: value)
.output flow_ea
.decl flow_aa(a1: value, a2: value)
.output flow_aa
.decl flow_ae(a1: value, e2: id)
.output flow_ae

.decl copy_ctx(from: context, to: context, e:id)
stored_val($VAddress(fv, to), v) :-
    copy_ctx(from, to, e),
    freevar(fv, e),
    stored_val($VAddress(fv, from), v).

state_e(eguard, ctx, $KAddress(eguard, ctx)),
stored_kont($KAddress(eguard, ctx), $If(et, ef, ctx, ak)),
flow_ee(e, eguard) :-
    state_e(e, ctx, ak),
    if(e, eguard, et, ef).

state_e(elam, ctx, $KAddress(elam, ctx)),
stored_kont($KAddress(elam, ctx), $Callcc(ectx, ak)),
flow_ee(e, elam) :-
    state_e(e, ctx, ak),
    callcc(e, elam),
    peek_ctx(e, ctx, ectx).

state_e(esetto, ctx, $KAddress(esetto, ctx)),
stored_kont($KAddress(esetto, ctx), $Set($VAddress(x, ctx), ak)),
flow_ee(e, esetto) :-
    state_e(e, ctx, ak),
    setb(e, x, esetto).

state_e(efunc, ctx, $KAddress(efunc, ctx)),
stored_kont($KAddress(efunc, ctx), $Arg(eargs, ctx, ectx, ak)),
flow_ee(e, efunc) :-
    state_e(e, ctx, ak),
    call(e, efunc, eargs),
    peek_ctx(e, ctx, ectx).

state_e(ebnd, ctx, $KAddress(ebnd, ctx)),
stored_kont($KAddress(ebnd, ctx), $Let($VAddress(x, ectx), ebody, ectx, ak)),
copy_ctx(ctx, ectx, e),
flow_ee(e, ebnd) :-
    state_e(e, ctx, ak),
    let(e, ll, ebody),
    let_list(ll, x, ebnd),
    peek_ctx(e, ctx, ectx).

state_e(earg0, ctx, $KAddress(earg0, ctx)),
stored_kont($KAddress(earg0, ctx), $Prim1(op, earg1, ctx, ak)),
flow_ee(e, earg0) :-
    state_e(e, ctx, ak),
    prim_call(e, op, pl),
    call_arg_list(pl, 0, earg0),
    call_arg_list(pl, 1, earg1).

state_a($Number(n), ak),
flow_ea(e, $Number(n)) :-
    state_e(e, ctx, ak),
    num(e, n).

state_a($Bool(b), ak),
flow_ea(e, $Bool(b)) :-
    state_e(e, ctx, ak),
    bool(e, b).

state_a($Closure(e, ctx), ak),
flow_ea(e, $Closure(e, ctx)) :-
    state_e(e, ctx, ak),
    lambda(e, _, _).

state_a(v, ak),
flow_ea(e, v) :-
    state_e(e, ctx, ak),
    var(e, x),
    stored_val($VAddress(x, ctx), v).

state_e(et, ctx_k, next_ak),
flow_ae($Bool("#t"), et) :-
    (state_a($Bool("#t"), ak) ; state_a($Closure(_,_), ak)
            ; state_a($Number(_), ak) ; state_a($Kont(_), ak)),
    stored_kont(ak, $If(et, _, ctx_k, next_ak)).

state_e(ef, ctx_k, next_ak),
flow_ae($Bool("#f"), ef) :-
    state_a($Bool("#f"), ak),
    stored_kont(ak, $If(_, ef, ctx_k, next_ak)).

state_e(ebody, ectx, next_ak),
stored_val($VAddress(x, ectx), $Kont(ak)),
copy_ctx(ctx_clo, ectx, elam),
flow_ae($Closure(elam, ctx_clo), ebody) :-
    state_a($Closure(elam, ctx_clo), ak),
    stored_kont(ak, $Callcc(ectx, next_ak)),
    lambda(elam, params, ebody),
    lambda_arg_list(params, 0, x).

state_a($Kont(ak), bk),
flow_aa($Kont(bk), $Kont(ak)) :-
    state_a($Kont(bk), ak),
    stored_kont(ak, $Callcc(_, _)).

state_e(earg, ctx, $KAddress(earg, ctx)),
stored_kont($KAddress(earg, ctx), $Fn(v, pos, ectx, next_ak)),
flow_ae(v, earg) :-
    state_a(v, ak),
    stored_kont(ak, $Arg(eargs, ctx, ectx, next_ak)),
    call_arg_list(eargs, pos, earg).

state_e(ebody, ectx, next_ak),
stored_val($VAddress(x, ectx), v),
copy_ctx(ctx_clo, ectx, elam),
flow_ae(v, ebody) :-
    state_a(v, ak),
    stored_kont(ak, $Fn($Closure(elam, ctx_clo), pos, ectx, next_ak)),
    lambda(elam, params, ebody),
    lambda_arg_list(params, pos, x).

state_a(v, callcc_kont),
flow_aa(v, v) :-
    state_a(v, ak),
    stored_kont(ak, $Fn($Kont(callcc_kont), 0, _, _)).

state_e(ebody, ctx, next_ak),
stored_val(av, v),
flow_ae(v, ebody) :-
    state_a(v, ak),
    stored_kont(ak, $Let(av, ebody, ctx, next_ak)).

state_e(earg1, ctx, $KAddress(earg1, ctx)),
stored_kont($KAddress(earg1, ctx), $Prim2(op, v, next_ak)),
flow_ae(v, earg1) :-
    state_a(v, ak),
    stored_kont(ak, $Prim1(op, earg1, ctx, next_ak)).

state_a($PrimVal(op, v1, v2), next_ak),
flow_aa(v2, $PrimVal(op, v1, v2)) :-
    state_a(v2, ak),
    stored_kont(ak, $Prim2(op, v1, next_ak)).

state_a($Number(-42), next_ak),
stored_val(loc, v),
flow_aa(v, $Number(-42)) :-
    state_a(v, ak),
    stored_kont(ak, $Set(loc, next_ak)).
\end{Verbatim}

\end{document}